\documentclass[reprint,
aip,apl, 
10pt,
superscriptaddress,
amsmath,amssymb,
floatfix
]{revtex4-1} 

\usepackage{amsmath,amssymb,bm}
\usepackage{mathtools}
\usepackage{esvect}
\usepackage{multirow}
\usepackage{array}
\usepackage[svgnames]{xcolor}
\usepackage[colorlinks=true,
linkcolor=MediumBlue,
urlcolor=MediumBlue,
citecolor=MediumBlue]{hyperref}
\begin{document}

\title{\LARGE Absolute pressure and gas species identification with an \\ optically levitated rotor}

\author{Charles P. Blakemore}
\email{cblakemo@stanford.edu}
\affiliation{Department of Physics, Stanford University, Stanford, California 94305, USA}

\author{Denzal Martin}
\affiliation{Department of Physics, Stanford University, Stanford, California 94305, USA}

\author{Alexander Fieguth}
\affiliation{Department of Physics, Stanford University, Stanford, California 94305, USA}

\author{Akio Kawasaki}
\affiliation{Department of Physics, Stanford University, Stanford, California 94305, USA}
\affiliation{W. W. Hansen Experimental Physics Laboratory, Stanford University, Stanford, California 94305, USA\looseness=-1}

\author{Nadav Priel}
\affiliation{Department of Physics, Stanford University, Stanford, California 94305, USA}

\author{\\Alexander D. Rider}
\thanks{Now at SRI International, Boulder, Colorado 80302}
\affiliation{Department of Physics, Stanford University, Stanford, California 94305, USA}

\author{Giorgio Gratta}
\affiliation{Department of Physics, Stanford University, Stanford, California 94305, USA}
\affiliation{W. W. Hansen Experimental Physics Laboratory, Stanford University, Stanford, California 94305, USA\looseness=-1}

\date{\today}
\begin{abstract}
\small
The authors describe a novel variety of spinning-rotor vacuum gauge in which the rotor is a ${\sim}4.7{\text -}\mu$m-diameter silica microsphere, optically levitated. A rotating electrostatic field is used to apply torque to the permanent electric dipole moment of the silica microsphere and control its rotational degrees of freedom. When released from a driving field, the microsphere's angular velocity decays exponentially with a damping time inversely proportional to the residual gas pressure, and dependent on gas composition. The gauge is calibrated by measuring the rotor mass with electrostatic co-levitation, and assuming a spherical shape, confirmed separately, and uniform density. The gauge is cross-checked against a capacitance manometer by observing the torsional drag due to a number of different gas species. The techniques presented can be used to perform absolute vacuum measurements localized in space, owing to the small dimensions of the microsphere and the ability to translate the optical trap in three dimensions, as well as measurements in magnetic field environments. In addition, the dynamics of the microsphere, paired with a calibrated vacuum gauge, can be used to measure the effective molecular mass of a gas mixture without the need for ionization and at pressures up to approximately 1~mbar. \\

\noindent DOI: \href{https://doi.org/10.1116/1.5139638}{10.1116/1.5139638}

\end{abstract}

\maketitle

\section{INTRODUCTION}
\vspace*{-0.3cm}

Vacuum technology plays an integral role in science and technology. While a number of different technologies exist,\cite{ohanlon,lesker,leybold,mks,pfeiffer} under high vacuum conditions, absolutely calibrated pressure measurements are challenging. Some of the most sensitive pressure gauges ionize residual gas molecules and measure the resulting electrical current. Such ionization gauges require an empirical calibration, accounting for the efficiency of the ionization technique employed, which varies across molecular species and filament materials, and may change over time.\cite{Lawrence:1926,Langmuir:1928,Gulyaev:1967} Additionally, the production of cations and their associated electrons, inherent to ionization gauges, can have a detrimental effect on conditions within an experimental chamber. At higher pressures, the measurement of heat transport in a gas does not require ionization but still requires empirical and species-dependent calibrations.

Gauges based on mechanical measurements and the kinetic theory of gases are known to provide absolute measurements of pressure. For example, capacitance manometers measure the force on a membrane exposed to the residual gas. Force sensitivity limits the pressure accessible to such gauges to approximately $10^{-5}~$mbar.\cite{mks} By comparison, spinning-rotor gauges measure the torsional drag induced by residual gas on a macroscopic rotor (an idea originally proposed by Maxwell~\cite{Maxwell:1866}) which, in existing devices, is generally magnetically levitated. At high vacuum, in the molecular flow regime, such drag can be simply related to the pressure,\cite{Beams:1962,Comsa:1980,McCulloh:1983,Cavalleri:2010,Kuhn:2017,Martinetz:2018} resulting in an absolute calibration. The minimum measurable pressure, approximately $5\times10^{-7}~$mbar, is usually limited by systematic uncertainties or the required integration times.\cite{McCulloh:1983,Fremerey:1985,Chang:2007} 

We have developed a spinning-rotor vacuum gauge based on an optically levitated, electrically driven microsphere (MS). The torsional drag on the MS can be measured by analyzing the rotational dynamics of the trapped MS. While silica MSs commonly used in optical levitation experiments\cite{Ashkin:1986,Li:2011,Gieseler:2012,Li:2013,Yin:2013,Moore:2014,Millen:2015,Rider:2016,Jain:2016,Ranjit:2016,Monteiro:2017,Hempston:2017,Rider:2018,Monteiro:2018,Ricci:2018,Blakemore:2019,Rider:2019,Blakemore:2019_2} can be electrically neutralized,\cite{Moore:2014,Rider:2016,Monteiro:2017,Rider:2018,Monteiro:2018,Ricci:2018,Blakemore:2019,Rider:2019,Blakemore:2019_2} they have been shown to have residual electric dipole moments.\cite{Moore:2014,Rider:2016} Rotation can then be induced by applying torque with an electric field while being measured optically, as described in Ref.~\onlinecite{Rider:2019}. Similar techniques with an electric drive and optical readout have been demonstrated for graphene nanoplatelets in a Paul trap.\cite{Coppock:2016}

In this scheme, the driving and interrogation mechanisms are not influenced by magnetic fields, and thus, the gauge can operate in conditions inaccessible to conventional spinning-rotor gauges. Another significant advantage of this technology, most applicable to experiments based on optically levitated particles, is that the measurement of pressure is relative to the environment in the immediate vicinity of the particle itself. A translation of the optical trap, or an array of optical traps each filled with a single rotating microsphere, would thus allow for mapping of pressure gradients with a spatial resolution limited only by the dimensions of the trapped particle(s).

\vspace*{-0.3cm}
\section{EXPERIMENTAL APPARATUS}
\vspace*{-0.3cm}

\begin{figure}[t!]
\includegraphics[width=1.0\columnwidth]{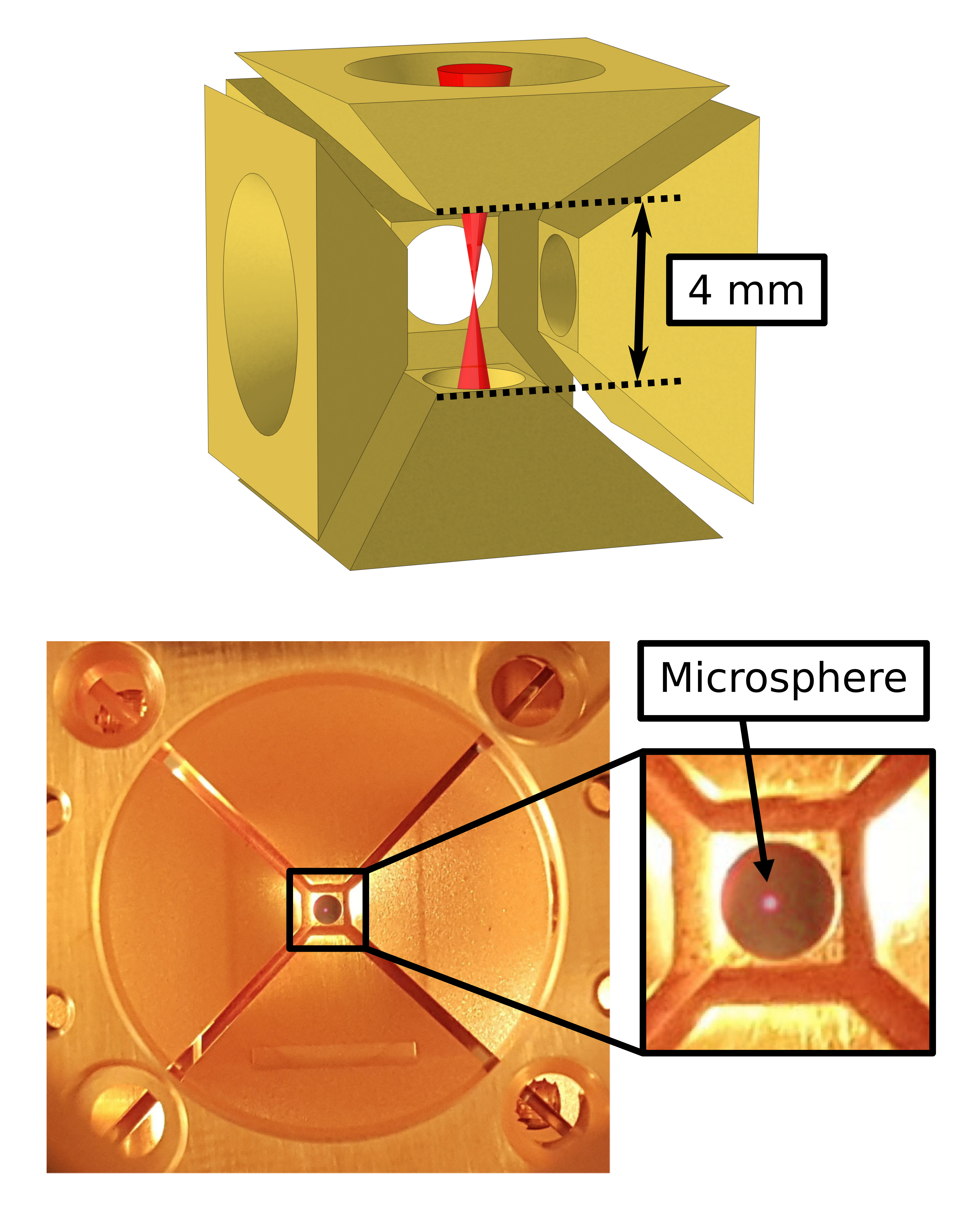}
\caption{\small A schematic of the inner region of the shielding electrodes with a photograph of the same structure and a trapped microsphere, visible through a quartz vacuum window. The foreground electrode has been removed from the schematic in order to clearly show the trapping beam, as well as from the actual apparatus in order to capture this photograph and show the trapped microsphere. The trapping beam shown has a numerical aperture of 0.12 and a Rayleigh range of ${\sim}25~\mu$m. For all measurements discussed, all six electrodes are in place.}
\label{fig:apparatus}
\end{figure}

The apparatus is identical to that presented in Refs.~\onlinecite{Blakemore:2019,Blakemore:2019_2,Rider:2019}. Briefly, a vertically oriented optical tweezer formed in vacuum by two identical aspheric lenses is surrounded by six independently biased electrodes, used to shield or manipulate the MS. A schematic of the inner region of the electrodes is shown in Fig.~\ref{fig:apparatus}, together with a photograph of a trapped microsphere within the same electrode structure. One electrode in the foreground of both the schematic and the photo has been removed to facilitate viewing. A field-programmable gate array serves as a multi-channel direct-digital-synthesis waveform generator which drives the six electrodes to produce arbitrary electric fields inside the trapping region. Importantly, this system is capable of generating electric fields that are constant in magnitude, but rotating in direction, in order to drive the MS's rotational degrees of freedom (DOFs) through their permanent electric dipole moment.

An optical system, detailed in Refs.~\onlinecite{Rider:2018,Blakemore:2019}, is used to measure the MS's three translational DOFs and provide stabilizing feedback under high vacuum conditions. Polarization-sensitive optics, described in Ref.~\onlinecite{Rider:2019}, measure the power of light in the polarization orthogonal to the trapping beam. Rotating, birefringent particles, such as the silica MSs used here, couple some linearly-polarized incident light into the orthogonal polarization at twice the frequency of their rotation, as demonstrated in Ref.~\onlinecite{Rider:2019}. The phase of this power modulation can be used to deduce the rotation angle of the MS.

The vacuum system is similar to that in Ref.~\onlinecite{Rider:2019} with the addition of a residual gas analyzer (RGA), and a manifold providing He, N$_2$, Ar, Kr, Xe, and SF$_6$ gases to the chamber. The manifold can be evacuated with a dedicated scroll pump. Capacitance manometers, a Pirani gauge, and the RGA, serving as an ionization gauge, together determine pressures in the range ${\sim}10^{-6}-10^3$~mbar and, with reduced accuracy, down to ${\sim}10^{-8}$~mbar. This system allows for control of the species and pressure of gas present in the experimental chamber.

\vspace*{-0.3cm}
\section{TORQUE NOISE AND TORSIONAL DRAG}
\vspace*{-0.3cm}

Spinning-rotor vacuum gauges operate in the molecular flow regime, where they make use of the proportionality between gas pressure and the torsional drag induced by the gas in order to measure the chamber pressure. From the fluctuation dissipation theorem, the torsional drag coefficient $\beta_{\rm rot}$ is proportional to the single-sided torque noise power spectral density $S_N$,

\vspace*{-0.3cm}{}
\begin{equation} \label{eq:fluc-diss}
\begin{split}
S_N &= 4 k_B T \beta_{\rm rot},
\end{split}
\end{equation}

\noindent where $k_B = 1.381 \times 10^{-23}~\rm{m}^2\,\rm{kg}\,\rm{s}^{-2}\,\rm{K}^{-1}$ is the Boltzmann constant and $T$ is the temperature of the residual gas. $S_N$ can also be computed by considering the force noise spectral density, $S_{F_\parallel}$, imparted parallel to a surface element of the MS by successive gas molecule collisions and adsorption, followed by thermalization and re-emission, the common way to treat diffusive scattering. $S_{F_\parallel}$ can be integrated across the surface of the MS to compute $S_N$, which is then combined with Eq.~(\ref{eq:fluc-diss}) to obtain the relation,

\vspace*{-0.3cm}{}
\begin{equation} \label{eq:rot-damping}
\begin{split}
\beta_{\rm rot} \times \frac{v_{\rm th}}{P} &= \pi r^4 \left( \frac{32}{9 \pi} \right)^{1/2},
\end{split}
\end{equation}

\noindent where $v_{\rm th} = \sqrt{k_B T / m_0}$ is the characteristic thermal velocity of residual gas molecules of mass $m_0$ at temperature $T$, $P$ is the gas pressure, and $r$ is the MS radius. This is identical to the results in Refs.~\onlinecite{Cavalleri:2010,Martinetz:2018}, and allows an absolute measurement of the pressure $P$ via $\beta_{\rm rot}$. The right hand side of Eq.~(\ref{eq:rot-damping}) is a geometric factor that depends only on $r$, the radius of the MS itself. Thus, if the gas species (i.e. $v_{\rm th}$) is known, a measurement of the torsional drag constant $\beta_{\rm rot}$ can be used to directly infer the gas pressure from Eq.~(\ref{eq:rot-damping}).

Two phenomena can be used to measure the torsional drag $\beta_{\rm rot}$ produced on a spinning MS. (1)~When released from a driving field, the angular momentum of the MS will decay exponentially with a time constant inversely proportional to the torsional drag coefficient. This is the primary method used to measure pressure with a spinning-rotor gauge and depends only on knowledge of the spherical rotor's radius and moment of inertia.\cite{Rider:2019} (2)~When driven with an electric field, the torque exerted on the MS is proportional to the sine of the angle between the rotating field and the MS's permanent dipole moment. Gas surrounding the MS produces a drag torque which determines the equilibrium angle between the driving field and the MS's dipole moment. This angle can be measured as a phase difference between the driving field and the power modulation produced in the cross-polarized light.\cite{Rider:2019}

It is distinctly possible that translational damping on resonance could also be used to infer the vacuum pressure. However, optically trapped MSs in the described apparatus exhibit instabilities at high vacuum, requiring active feedback in order to retain them for long periods~\cite{Moore:2014,Rider:2016,Ranjit:2016,Rider:2018,Blakemore:2019,Rider:2019,Blakemore:2019_2}. The active feedback is designed primarily with derivative gain to mimic the damping provided by residual gas in lower vacuum environments where the trap is stable. It is difficult to deconvolve this feedback with the damping provided by actual residual gas in the vicinity of the MS. At first order, the rotational degrees of freedom are unaffected by the stabilizing feedback.

\vspace*{-0.3cm}
\section{SPINDOWN PRESSURE MEASUREMENTS}
\vspace*{-0.3cm}

From Eq.~(\ref{eq:rot-damping}), the pressure can be written in terms of a constant $\kappa=\kappa(T, r)$,  the gas particle mass $m_0$, and the torsional drag coefficient $\beta_{\rm rot}$,

\vspace*{-0.3cm}
\begin{equation} \label{eq:pressure_meas}
\begin{split}
P &= \frac{\kappa}{\sqrt{m_0}} \cdot \beta_{\rm rot},
\end{split}
\end{equation}

\noindent where $\kappa \equiv (1/r^4) \sqrt{9 k_B T / 32 \pi}$ is determined by measuring the MS radius $r$ and the experimental chamber temperature $T$. The MS radius is determined indirectly by measuring the mass of the MS via electrostatic co-levitation,\cite{Blakemore:2019_2} and assuming the same uniform density found in,\cite{Blakemore:2019_2} as the MSs are derived from the same lot. The drag coefficient $\beta_{\rm rot}$ is measured by observing the spindown time of the MS released from a driving field. If torsional drag due to residual gas is the only torque present, then the angular velocity of the MS will decay exponentially with time constant $\tau = I / \beta_{\rm rot}$, where $I$ is the MS moment of inertia.

In fact, there is a roughly constant residual torque, likely optical in nature and generated from a small ellipticity in the trapping beam, which couples to the residual birefringence of the MS, as discussed in Refs.~\onlinecite{Monteiro:2018,Rider:2019}. Under these conditions, the equation of motion for the MS angular momentum and the solution for the frequency of rotation are given by:

\vspace*{-0.3cm}
\begin{gather} 
\frac{d L}{d t} = N_{\rm opt} - \frac{\beta}{I} L(t), \label{eq:eom} \\
\implies f(t) = f_0 e^{- (t - t_0) / \tau} + f_{\rm opt} (1 - e^{- (t - t_0) / \tau}), \label{eq:spindown}
\end{gather}

\noindent where $L(t)$ is the angular momentum of the MS, $f_0$ is the initial rotation frequency at time $t=t_0$, $\tau = I / \beta_{\rm rot}$ is the decay time due to gas drag, and $f_{\rm opt} = N_{\rm opt} / (2 \pi \beta_{\rm rot})$ is the terminal rotation frequency of the MS due to a constant optical torque $N_{\rm opt}$. The quantity $f_{\rm opt}$ can be measured by allowing the MS to reach steady state in the absence of any electrical driving torques and observing the terminal rotation frequency. Three different MSs were used in this work, with generally consistent results. For MS~No.~1, the terminal rotation frequency was measured to be $f_{\rm opt} = (8315 \pm 559)~$Hz, where the uncertainty is the maximum observed deviation from the central value over approximately 6~hrs of successive measurements, where for each 2-s integration, the terminal rotation can be measured to within $\pm1$~Hz uncertainty. This type of systematic effect is similar to the ``offset correction'' necessary for magnetically levitated spinning-rotor gauges.\cite{Comsa:1980,McCulloh:1983,Fremerey:1985,Chang:2007}

To measure $\tau$, the MS is released from an $f_0 = 110~$kHz driving field, and the frequency of power modulation of cross-polarized light produced by the rotation of the MS is observed. Successive 2-s integrations, separated by approximately one second, are first bandpass-filtered and then Hilbert-transformed to recover the instantaneous frequency of power modulation as a function of time. The median time of the $i$-th integration, $\langle t_i \rangle$, relative to the field turning off at $t_0$, and the arithmetic mean of the instantaneous frequency measured during this integration, $\langle f_i \rangle$, are used to calculate a value for $\tau$ from Eq.~(\ref{eq:spindown}):

\vspace*{-0.3cm}
\begin{equation}
\tau_i = \frac{\langle t_i \rangle - t_0}{\log \left( \frac{f_0 - f_{\rm opt}}{\langle f_i \rangle - f_{\rm opt}} \right)}. \label{eq:tau}
\end{equation}

\begin{figure}[b!]
\includegraphics[width=1.0\columnwidth]{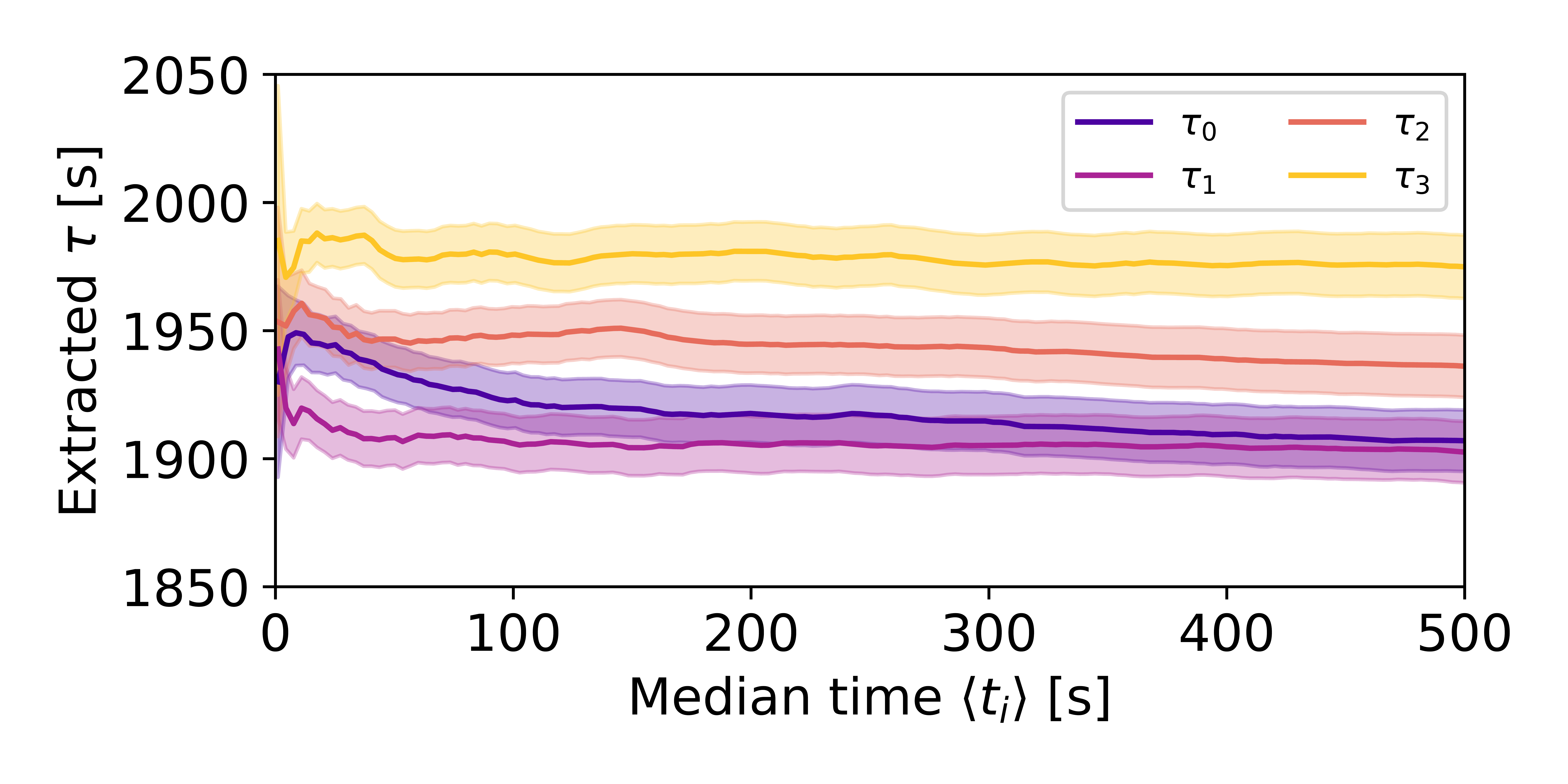}
\caption{\small Damping time $\tau$ calculated from Eq.~(\ref{eq:tau}) versus median integration time $\langle t_i \rangle$, for the four measurements with MS~No.~1, indexed chronologically. The bands show the 1$\sigma$ uncertainty propagated from the individual uncertainties of the values used to calculate $\tau$.}
\label{fig:spindown_time}
\end{figure}

The calculated value of $\tau_i$ is plotted as a function of median integration time in Fig.~\ref{fig:spindown_time} for four example measurements performed with MS~No.~1. The shaded bands represent the quadrature sum of systematic and statistical uncertainties, and are dominated by a systematic uncertainty in the assumed value of the terminal rotation frequency. For a single measurement, the estimation of $\tau$ from Eq.~(\ref{eq:tau}) is self-consistent for approximately the first 500-s. For longer times, the effect of a slowly changing optical torque skews an estimation of $\tau$ by up to $10\%$.

\begin{figure}[b!]
\includegraphics[width=1.0\columnwidth]{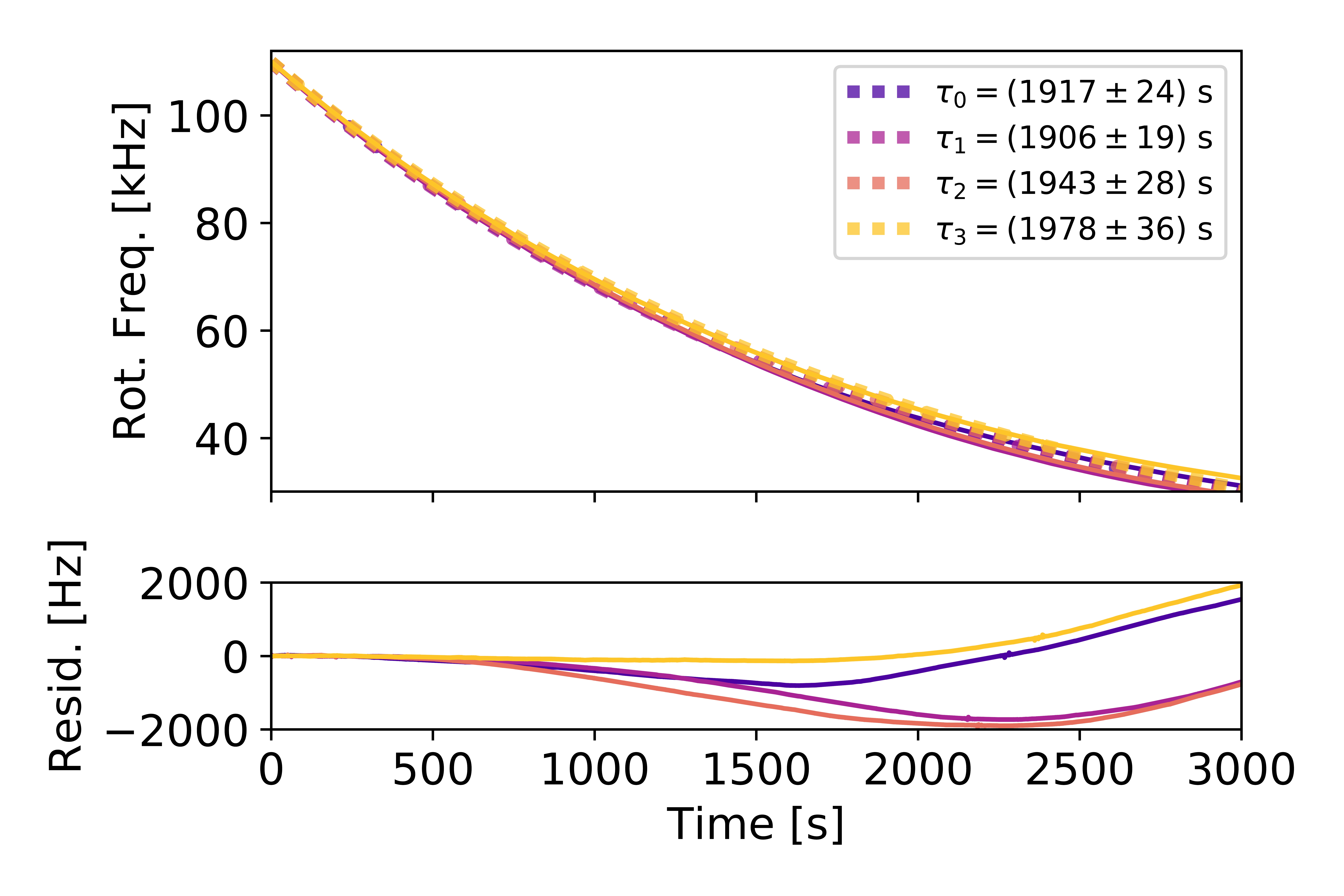}
\caption{\small Exponential decay of MS~No.~1 angular velocity due to torsional drag from residual gas after a 110~kHz driving field has been turned off. Data are shown for the same four measurements detailed in Fig.~\ref{fig:spindown_time}. Dotted lines indicated realizations of Eq.~(\ref{eq:spindown}) with $\tau = \tau_j$ for measurements $j=0,1,2,3$, where $\tau_j$ is the average of the values plotted in Fig.~\ref{fig:spindown_time}. The structure in the residuals, with amplitude ${\sim}2\%$, is likely the result of a slowly fluctuating optical torque.}
\label{fig:spindown}
\end{figure}

The exponential decay of the rotation frequency for each of the four measurements with MS~No.~1 are detailed in Fig.~\ref{fig:spindown}, where data are shown for about $1~$hour following the release of MS~No.~1. Data in Fig.~\ref{fig:spindown} are overlaid with Eq.~(\ref{eq:spindown}), where the values of $\tau_j$ for measurements $j=0,1,2,3$ are the mean values of $\tau_i$ for $\langle t_i \rangle < 500~$s, and $f_{\rm opt} = 8315~$Hz is fixed. The structure in the residuals, with an amplitude approximately $2\%$ of $f_0$, is likely the result of the slowly fluctuating optical torque.

The values of $P$ computed with Eq.~(\ref{eq:pressure_meas}) are shown in Table~\ref{table:pmeas}, assuming $m_0 = 18~$amu, as RGA analysis found the residual pressure to be dominated by water vapor. The four successive measurements, taken over the course of a few days with the same MS, are consistent within statistical uncertainties, and the systematic uncertainty should be common to all. Magnetically levitated spinning-rotor gauges typically have similar measurement uncertainties.\cite{Beams:1962,Fremerey:1985,Berg:2015} As a specific example, at a calibration pressure of $2.0\times10^{-4}~$mbar, NIST calibration services for steel spinning-rotor gauges report a 0.02\% relative uncertainty,\cite{Berg:2015} i.e., $\sigma_P \approx 4\times10^{-8}~$mbar. This is within a factor of five of the uncertainty of our newly demonstrated silica MS spinning-rotor gauge (Table~\ref{table:pmeas}).

\begin{table}[t!]
  \caption{ \small Results of spindown pressure measurements, taken with MS~No.~1. Statistical and systematic uncertainties are propagated independently from uncertainties on $r$, $I$, and $\tau$.}
  \footnotesize
  \label{table:pmeas}
  \vspace*{1em}
  \setlength{\tabcolsep}{2em}
  {\renewcommand{\arraystretch}{1.4}
  \begin{tabular}{cc}
    \toprule
    \, & \, \small{$P$\quad[$10^{-6}$~mbar]} \\
    \colrule
    $P_0$ & $3.53 \pm 0.17{\rm\,(stat.)} \pm 0.30 {\rm\,(sys.)}$ \\
    $P_1$ & $3.56 \pm 0.17{\rm\,(stat.)} \pm 0.31 {\rm\,(sys.)}$  \\
    $P_2$ & $3.49 \pm 0.17{\rm\,(stat.)} \pm 0.30 {\rm\,(sys.)}$  \\
    $P_3$ & $3.43 \pm 0.16{\rm\,(stat.)} \pm 0.29 {\rm\,(sys.)}$  \\
    \botrule
  \end{tabular}}
\end{table}

Importantly, these are measurements of the chamber pressure in the immediate vicinity of the MS, so that they can be used to estimate force and torque noise in precision measurement applications. The total pressure measured by the RGA is lower by approximately a factor of three, which could be attributed to a poor filament or response calibration, or may be indicative of a real pressure difference across the experimental chamber. With the current apparatus, there is a substantial pumping impedance between the trapping region and the RGA filament, as the trap is confined within the six electrodes which form a cage.

While the presence of anomalous sources of dissipation can't be excluded, two possible sources are shown to be too small to account for the pressure difference observed. The image charge induced on the 6 electrodes by the electric dipole moment would exert a torque roughly six orders of magnitudes smaller than the drag torque at the initial rotation velocity. Anomalous damping may also arise from electric field noise on the, nominally grounded, driving electrodes. Results in plasma physics suggest that fluctuating fields in the frequency band around the rotation frequency of the MS could also tend to increase the angular velocity of the MS through stochastic acceleration,\cite{Sturrock:1966} resulting in an apparent pressure that would be lower instead of higher.

To test the second phenomenon, the equation of motion given by Eq.~(\ref{eq:eom}) was numerically integrated, including 100 distinct implementations of random torque fluctuations given by Eq.~(\ref{eq:fluc-diss}), and distinct realizations of anomalous torque from the electric field noise produced by the driving electronics when the field is nominally off. As this anomalous torque is incoherent with the decaying angular momentum of the MS, its effect on the apparent damping time was found to be less than 0.1\%, and can't explain the higher value of the measured pressure.

The apparent pressure difference may also be due to an elevated temperature of the MS, which is a notoriously difficult quantity to measure,\cite{Millen:2014,Arita:2016,Monteiro:2017,Hebestreit:2018,Delic:2019} often yielding only an upper bound. The derivation of Eq.~(\ref{eq:rot-damping}) assumes that the MS itself is in thermal equilibrium with the surrounding gas, so that incident and re-emitted gas particles have the same Maxwellian thermal velocity. If this assumption is removed, then the right-hand side of Eq.~(\ref{eq:pressure_meas}) would have a multiplicative factor $[ 2 T_{\rm gas} / (T_{\rm gas} + T_{\rm MS})]$, with $T_{\rm gas}$ being the temperature of incident residual gas particles assumed to be in equilibrium with the chamber itself, and $T_{\rm MS}$ being the elevated temperature of the MS assumed to be in equilibrium with outgoing residual gas molecules. If the apparent pressure difference between the RGA and the location of the trap is due entirely to an elevated MS temperature, this implies $T_{\rm MS} \sim 1500~$K.

\vspace*{-0.3cm}
\section{EQUILIBRIUM PHASE LAG}
\vspace*{-0.3cm}

The second method proposed to measure $\beta_{\rm rot}$ can be substantially faster, but in our apparatus has only sensitivity for moderate vacuum. Since this is also the regime in which absolutely calibrated capacitance manometers operate, we use this method for a cross-check of the technique. Consider a MS rotating at fixed frequency under the influence of a rotating electric field. In the rotating reference frame, the equation of motion for the phase $\phi$ between the driving field and the orientation of dipole moment has an equilibrium solution developed in Ref.~\onlinecite{Rider:2019} and given by,

\vspace*{-0.3cm}{}
\begin{equation} \label{eq:phi-eq}
\begin{split}
\phi_{\rm eq} &= -{\rm arcsin} \, \left( \frac{\beta_{\rm rot} \omega_0}{E d} \right) = -{\rm arcsin} \, \left( \frac{P}{P_{\rm max}} \right), 
\end{split}
\end{equation}

\noindent where $\omega_0$ is the angular driving frequency, $E$ is the electric field magnitude, and $d$ is the permanent electric dipole moment of the MS. The expression is written in terms of a single parameter $P_{\rm max}$, the maximum pressure that has a valid equilibrium solution given particular values of $E$, $d$, and $\omega_0$. Above this pressure, the driving field can no longer provide sufficient torque to maintain the MS's rotation. Using Eq.~(\ref{eq:rot-damping}), 

\vspace*{-0.3cm}{}
\begin{equation} \label{eq:pmax}
\begin{split}
P_{\rm max} &= \frac{E d}{\omega_0} \frac{P}{\beta_{\rm rot}} \equiv \frac{E d}{\omega_0} \frac{\kappa}{\sqrt{m_0}},
\end{split}
\end{equation}

\noindent where $\kappa$ is already defined. The prefactor $(E d / \omega_0)$ is related to the driving torque and can be set by the experimenter, whereas $\kappa$ is a constant across gas species, as the MS radius is unchanged throughout the measurements.

\begin{figure}[b!]
\includegraphics[width=1.0\columnwidth]{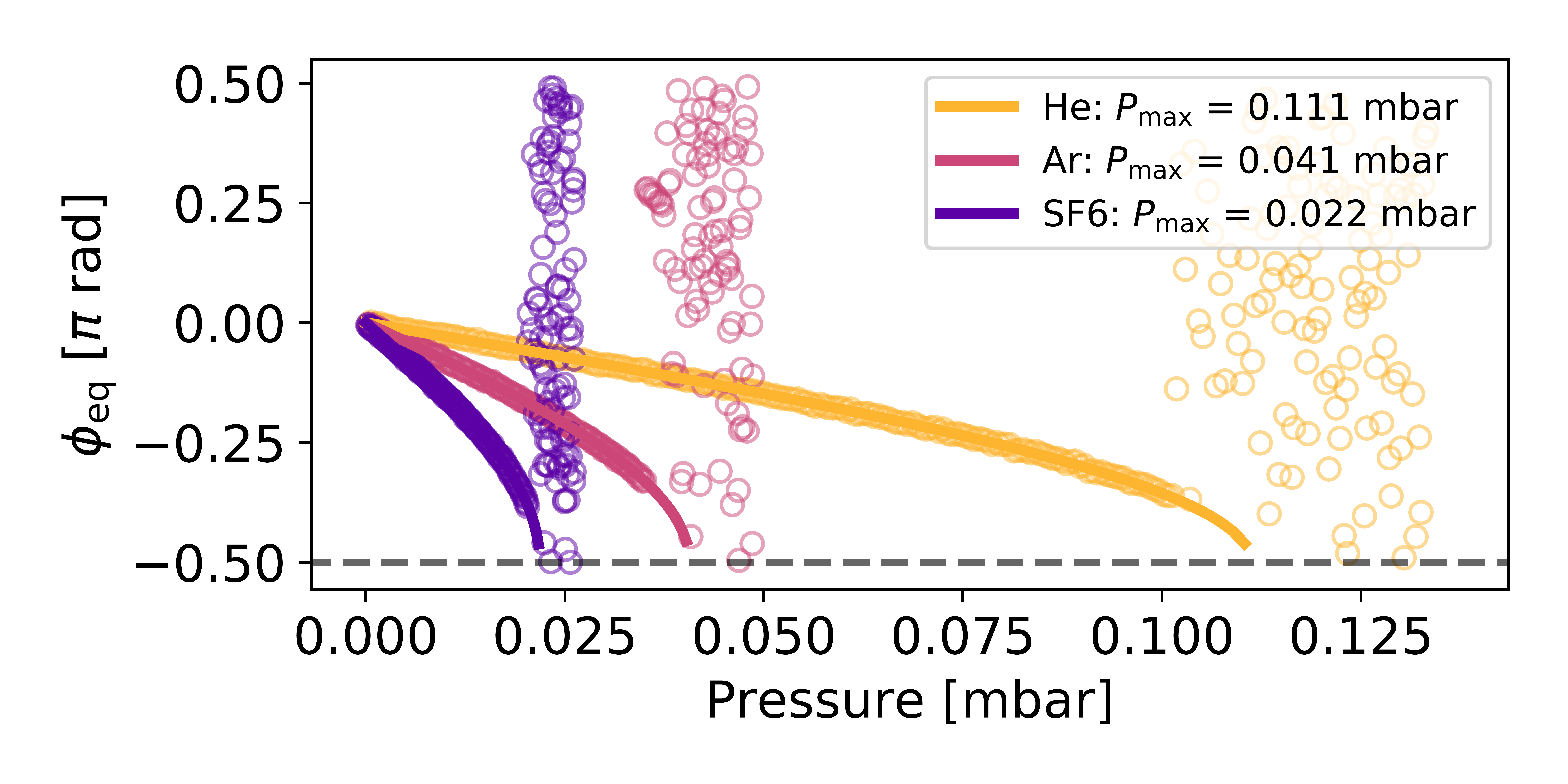}
\caption{\small Equilibrium phase lag between the orientation of the MS dipole moment and a rotating electric field, versus the residual gas pressure determined by two capacitance manometers. Three distinct measurements are shown with He, Ar, and SF$_6$ residual gas. As the pressure is increased, the phase lags according to Eq.~(\ref{eq:phi-eq}), and $P_{\rm max}$ is measured from a fit to this expression. As $\phi_{\rm eq}$ approaches $-\pi/2$, the MS rotation becomes unlocked from the driving field ($\phi$ becomes random) at a pressure slightly below $P_{\rm max}$, due to librational motion of the MS and torque fluctuations from the increasing pressure  of the surrounding gas.}
\label{fig:example_pramp}
\end{figure}

Thus, a gauge cross-check consists of measuring $P_{\rm max}$, which is inversely proportional to the damping constant, as a function of $m_0$, the particle mass of a residual gas species, while maintaining a constant driving torque and chamber temperature. $P_{\rm max}$ can be measured by linearly ramping the pressure, monitored with a capacitance manometer, while continuously measuring $\phi_{\rm eq}$ until unlocking, at which point the phase lag $\phi$ becomes random, as shown in Ref.~\onlinecite{Rider:2019}. 

This effect is demonstrated for He, Ar, and SF$_6$ in Fig.~\ref{fig:example_pramp}. Due to both the librational motion of the MS and torque noise from the increasing pressure of residual gas, the MS rotation becomes unlocked from the driving field at a pressure slightly below $P_{\rm max}$. Hence, $P_{\rm max}$ is determined by extrapolating the endpoint of the arcsine relationship in Eq.~(\ref{eq:phi-eq}), as seen in Fig.~\ref{fig:example_pramp}.

The measurement of $\phi_{\rm eq}$ is used to validate Eq.~(\ref{eq:rot-damping}) and cross-check the MS as a spinning-rotor gauge against the capacitance manometer. By comparison, magnetically levitated spinning-rotors often make use of static expansion (wherein a defined amount of gas is allowed to fill a chamber of known volume) in order to provide cross-checks with ion gauges. Practical limitations of the current apparatus, primarily the bandwidth of the driving electronics, place a lower bound on the pressure observable with this method at ${\sim}10^{-5}~$mbar, as at lower pressures $\phi_{\rm eq}$ is too small to measure. Rotation velocities greater than the ${\sim}100~$kHz achievable in the current system may allow for the extension of the method to lower pressure.

Silicon nanorods in counterpropagating beam traps have also been used to measure equilibrium phase lag induced by gas drag, but with the rotation driven optically by modulating the polarization (from linear to circular) of the trapping beams at fixed frequency.\cite{Kuhn:2017} Phase lag between the optical drive and the induced rotation of the silicon rotor was demonstrated to accurately predict gas pressures in the range $4-10$~mbar.

\vspace*{-0.3cm}
\section{GAS COMPOSITION AND CHANGES IN DIPOLE MOMENT}
\vspace*{-0.3cm}

In many vacuum chambers, the ultimate base pressure achieved is limited by a single residual gas species, such as water vapor or hydrogen, and Eq.~(\ref{eq:pmax}) is directly applicable. In this work, requiring the deliberate introduction of various gas species, a small correction needs to be applied to account for contamination, primarily by water. When there are multiple species, each contributes independently to the total torsional drag. Appealing to the analysis in Refs.~\onlinecite{Cavalleri:2010,Martinetz:2018}, we can add the torque noise spectra of multiple gases, each with partial pressure $P_i = \chi_i P_{\rm tot}$ and particle mass $m_{0,i}$,

\vspace*{-0.3cm}{}
\begin{equation} \label{eq:torque-noise-eff}
\begin{split}
S_{N, {\rm tot}} &= 4 k_B T \beta_{\rm rot} = 4 k_B T \left( \frac{P_{\rm tot}}{\kappa} \sqrt{m_{0, {\rm eff}}} \right), 
\end{split}
\end{equation}

\noindent where we have written the expression in terms of an effective mass, $m_{0, {\rm eff}} = \left( \sum_i \chi_i \sqrt{m_{0,i}} \right)^2$. The mole fractions, $\chi_i$, can be measured with an RGA, appropriately accounting for differences in ionization probability for different gas species. For He, N$_2$, Ar, and SF$_6$, the preparation of the manifold results in an apparent purity of ${\gtrsim}99.9\%$, limited by systematic uncertainties in RGA ionization probabilities for different gas species. In the cases of Kr and Xe, the apparent purity is ${\sim}99\%$, limited by water contamination in the manifold.

The process of characterizing the residual gas leaked into the experimental chamber makes use of an RGA on the experimental chamber itself, which tends to charge the trapped MS with an excess of electrons. The MS is returned to zero charge, prior to spinning up and performing the drag measurements, by repeated exposure to ultraviolet photons from an Xe flashlamp.\cite{Moore:2014,Rider:2016} The process of charging and discharging appears to change the MS electric dipole moment.

In order to measure $\kappa$ as defined in Eq.~(\ref{eq:pmax}), the term $(Ed/\omega_0)$ must be measured precisely. The quantities $E$ and $\omega_0$ are controlled and measured precisely, and via methods presented in Ref.~\onlinecite{Rider:2019}, it is possible to characterize the MS permanent dipole moment $d$. This is done by analyzing the librational motion about the instantaneous direction of the electric field, together with a calculation of moment of inertia from the measured mass of the MS. 

Over the course of all measurements presented here, the dipole moment assumed values between approximately $95$ and $120~e{\cdot}\mu$m for MS~No.~1, between $30$ and $55~e{\cdot}\mu$m for MS~No.~2, and between $80$ and $110~e{\cdot}\mu$m for MS~No.~3. Without the RGA filament on, the dipole moment was measured to be constant, within measurement uncertainties, over the course of a day.

\vspace*{-0.3cm}
\section{CAPACITANCE MANOMETER CROSS-CHECK}
\vspace*{-0.3cm}

\begin{figure}[t!]
\includegraphics[width=1.0\columnwidth]{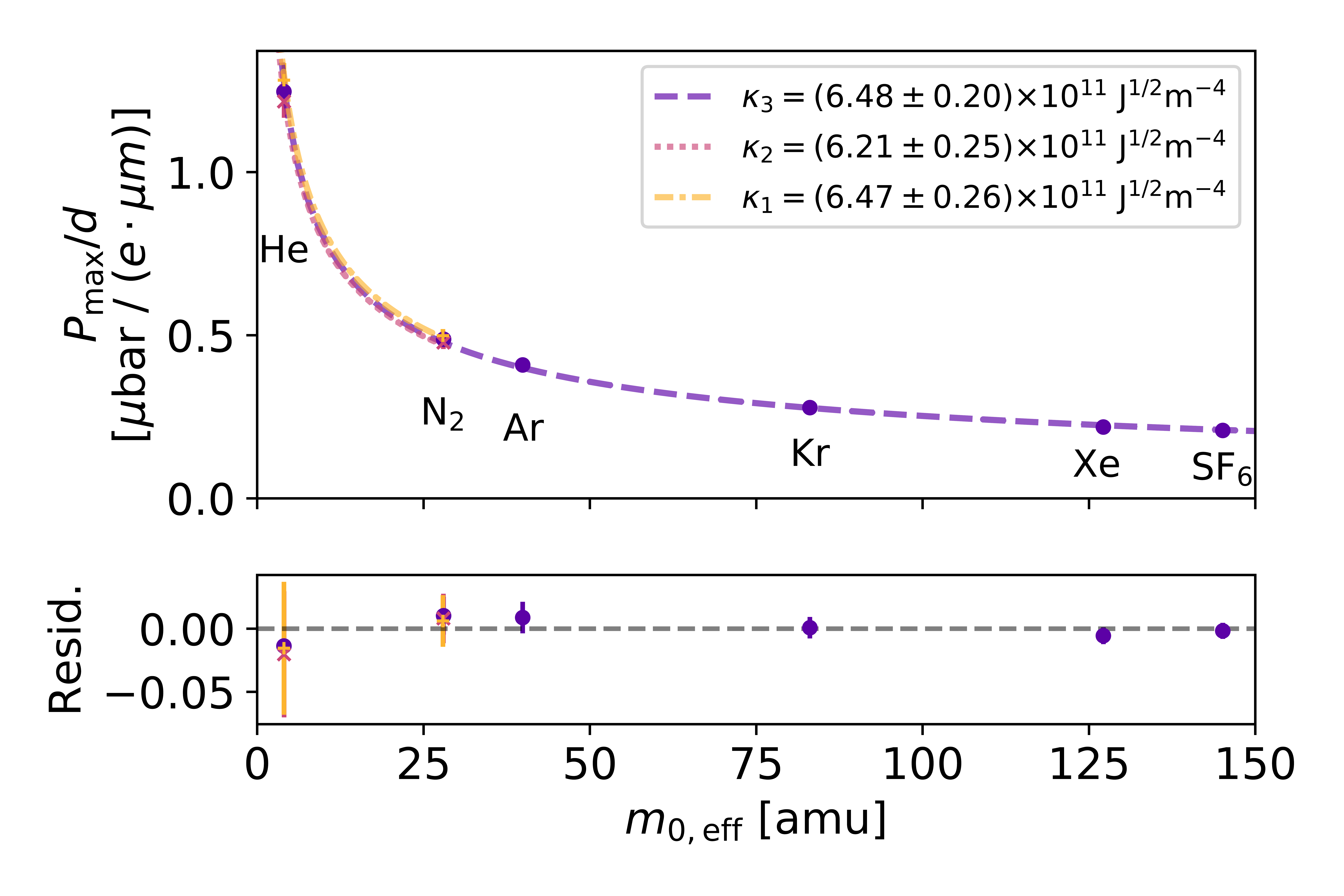}
\caption{\small The quantity $P_{\rm max}/d$ is plotted against the effective mass of the gas $m_{0,{\rm eff}}$ for different species spanning the 4 to 150~amu range, for three different MSs. With $E$ and $\omega_0$ known, the constant $\kappa$ is extracted by fitting Eq.~(\ref{eq:pmax}) to the data, with a single free parameter, as shown for each MS. A $\chi^2$-minimization finds $\chi_{1,{\rm min}}^2 / N_{\rm DOF} = 1.9/1$, $\chi_{2,{\rm min}}^2 / N_{\rm DOF} = 3.5/1$, $\chi_{3,{\rm min}}^2 / N_{\rm DOF} = 8.5/5$. Residuals shown below are plotted with the same units as the data.}
\label{fig:all_pramp}
\end{figure}

The MS spinning-rotor vacuum gauge is cross-checked via the following procedure: first, the gas manifold is prepared with a particular species and $m_{0,{\rm eff}}$ is characterized by leaking a sample of the gas into the experimental chamber where the RGA is present; second, the chamber is pumped to its base pressure, the MS is returned to a neutral state, and its permanent electric dipole moment $d$ is determined by analyzing the libration. Finally, the gas prepared is leaked into the experimental chamber, and the equilibrium phase lag $\phi_{\rm eq}$ is measured as a function of gas pressure to determine $P_{\rm max}$. The RGA filament is turned off prior to the second and third steps.

For MS~No.~1 and MS~No.~2, this procedure was performed three times for He and N$_2$, while for MS~No.~3 the measurement was performed with He, N$_2$, Ar, Kr, Xe, and SF$_6$. Successive measurements with each MS and gas species are found to agree and are averaged together by considering the quantity $(P_{\rm max} / d)$, to account for the small differences in dipole moment between measurements. The results of all pressure ramp measurements are shown in Fig.~\ref{fig:all_pramp}. A $\chi^2$-minimization with a single parameter is used to fit Eq.~(\ref{eq:pmax}) to the data and extract $\kappa$ for each MS. The data are well modeled by Eq.~(\ref{eq:pmax}), demonstrating the validity of Eq.~(\ref{eq:rot-damping}) and the analysis in Refs.~\onlinecite{Cavalleri:2010,Martinetz:2018}.

{The values of $\kappa$ determined with these analyses are shown in Table~\ref{table:kappa}, together with the individual MS radii and the expected value of $\kappa$, which have both been computed from the known value of the density~\cite{Blakemore:2019_2} and the measured value of the MS mass, assuming the MS is thermal equilibrium with the gas. The consistency between the measured and calculated values suggest that in moderate vacuum, $P \approx 10^{-3} - 10^{-1}~$mbar, the MS is indeed  \unskip\parfillskip 0pt \par} 

\onecolumngrid 

\begin{table}[b]
  \caption{ \small Calculated and measured calibration factors $\kappa$ for the three MSs used here, together with their measured masses, from which the values of $r$ and then $\kappa$ are determined. The MS density was assumed to be $\rho_{\rm MS} = 1.55\pm0.05{\rm\,(stat.)}\pm0.08{\rm\,(sys.)}$ from Ref.~\onlinecite{Blakemore:2019_2}. The relatively large uncertainty in $\rho_{\rm MS}$ limits the precision with which $\kappa$ can be calculated, whereas the precision of the measurement of $\kappa$ is limited by uncertainties in the measured dipole moment and $P_{\rm max}$.}
  \footnotesize
  \label{table:kappa}
  \vspace*{1em}
  \setlength{\tabcolsep}{2.25em}
  {\renewcommand{\arraystretch}{1.4}
  \begin{tabular}{lccc}
    \toprule
     & & \multicolumn{2}{c}{$\kappa$\quad[$10^{11}$ J$^{1/2}$m$^{-4}$]} \\\cline{3-4}
    \small{MS} & \small{$m_{\rm MS}$\quad[pg]} & \small{Theory} & \small{Experiment} \\
    \colrule
    No.~1 & $84.3 \pm 0.2{\rm\,(stat.)} \pm 1.5 {\rm\,(sys.)}$ & $6.3 \pm 0.3{\rm\,(stat.)} \pm 0.5 {\rm\,(sys.)}$ & $6.47 \pm 0.06{\rm\,(stat.)} \pm 0.25 {\rm\,(sys.)}$ \\
    No.~2 & $84.2 \pm 0.4{\rm\,(stat.)} \pm 1.4 {\rm\,(sys.)}$ & $6.3 \pm 0.3{\rm\,(stat.)} \pm 0.5 {\rm\,(sys.)}$ & $6.21 \pm 0.06{\rm\,(stat.)} \pm 0.24 {\rm\,(sys.)}$ \\
    No.~3 & $85.0 \pm 0.6{\rm\,(stat.)} \pm 1.5 {\rm\,(sys.)}$ & $6.2 \pm 0.3{\rm\,(stat.)} \pm 0.4 {\rm\,(sys.)}$ & $6.48 \pm 0.09{\rm\,(stat.)} \pm 0.17 {\rm\,(sys.)}$ \\
    \botrule
  \end{tabular}}
\end{table}
\twocolumngrid

\quad \\ 

\noindent in thermal equilibrium with the gas. The uncertainty in the directly measured value of $\kappa$ is dominated by systematic uncertainties in the measured dipole moment, which, in turn, depends on the uncertainties in the moment of inertia, another derived quantity.

Previous work with magnetically levitated spinning-rotor vacuum gauges usually include a momentum accommodation coefficient relating specular and diffuse reflection of gas particles, and which encompasses rotor geometry and surface composition. Extensive measurements have found this factor to be consistent with unity for steel rotors, with percent-level precision.\cite{Beams:1962,Comsa:1980,McCulloh:1983,Fremerey:1985,Dittmann:1989,Chang:2007} In this work, the accommodation coefficient would appear as a multiplicative constant $\sigma$ on the right-hand side of Eq.~(\ref{eq:rot-damping}). Comparing the measured and calculated values of $\kappa$, the accommodation coefficients $\sigma_i$ for the $i$-th MS were found to be consistent with unity: $\sigma_1 = 0.98 \pm 0.04{\rm\,(stat.)} \pm 0.08 {\rm\,(sys.)}$, $\sigma_2 = 1.02 \pm 0.05{\rm\,(stat.)} \pm 0.08 {\rm\,(sys.)}$, and $\sigma_3 = 0.96 \pm 0.04{\rm\,(stat.)} \pm 0.07 {\rm\,(sys.)}$. Prior work with a single levitated silica nanoparticle, with diameter $d\approx70~$nm, has found an accommodation coefficient of $\sigma = 0.65\pm0.08$~\cite{Hebestreit:2018} based on an analysis of heating rates. The discrepancy could be the result of the vastly different scale of the two types of rotors, poorly understood material properties, as well as a tendency for nanoscale particles to be highly non-spherical (see scanning electron microscopy images in Refs.~\onlinecite{Bishop:2004,Rahman:2012,Asenbaum:2013,Hoang:2016,Lu:2018,Dai:2018}).

The consistency between the measured and computed values of $\kappa$ is also an indirect validation of the work presented in Ref.~\onlinecite{Blakemore:2019_2} which makes no assumptions about MS temperature or thermal equilibrium and demonstrates independence across vacuum pressures $P \approx 10^{-6} - 10^{0}~$mbar. If the density computed there was incorrect, we would expect quantities derived from this density, such as $\kappa$, to be inconsistent with independent measurements of those quantities.

\vspace*{-0.3cm}
\section{NON-IONIZING GAS ANALYZER}
\vspace*{-0.3cm}

If the rotational dynamics of a MS with a known value of $\kappa$ are analyzed while the pressure is measured with a calibrated, species-independent vacuum gauge such as a capacitance manometer, the combination can be used as a non-ionizing gas analyzer, operable directly in moderate vacuum. In particular, this system would excel as a binary gas analyzer, comparing the concentrations of two gases, such as one might encounter in nanofabrication with dopants in a carrier gas, or in chemistry with a particular stoichiometric ratio of reagent gases. 

From Eq.~(\ref{eq:pressure_meas}), we can solve for the effective residual gas particle mass to find,

\vspace*{-0.3cm}{}
\begin{equation} \label{eq:gas_analyzer}
\begin{split}
m_{0,{\rm eff}} = \left( \frac{\kappa \beta_{\rm rot}}{P} \right)^2 = \left( \sum_i \chi_i \sqrt{m_{0,i}} \right)^2, 
\end{split}
\end{equation}

\noindent where $\beta_{\rm rot}$ can be determined from a spindown measurement, $\beta_{\rm rot} = I / \tau$, or from an equilibrium phase lag measurement, $\beta_{\rm rot} = E d \sin (-\phi_{\rm eq}) / \omega_0$, where $\phi_{\rm eq}$ is strictly negative. The latter quantity can be measured continuously. Such a system has an immediate advantage over ionizing RGAs, as it can operate directly in the pressure regime $P=10^{-5} - 10^{0}~$mbar, whereas an ionizing RGA requires $P < 10^{-4}~$mbar and is often connected to a vacuum chamber of interest by a leak valve, which can itself change the relative concentrations of gases. Indeed, the total absence of ionization offers its own advantage.

\vspace*{-0.3cm}
\section{CONCLUSION}
\vspace*{-0.3cm}

We have demonstrated the use of an optically levitated silica microsphere as a microscopic spinning-rotor vacuum gauge. The gauge operates on the principle that the torsional drag is proportional to the gas pressure in the vicinity of the microsphere. A spinning microsphere driven by a rotating electric field is released from the driving field, and the subsequent decay of angular momentum induced by torsional gas drag is observed. The residual gas pressure is then inferred from the decay time and the calculated moment of inertia. 

Once a microsphere has been calibrated by measuring its mass, pressures within the range $10^{-6} - 10^{-3}~$mbar can be determined within 10~s with a precision of ${\lesssim}7\%$, under the given assumptions and limited by the uncertainty in the calibration $\kappa$, which, in turn, depends on uncertainty in the assumed microsphere density. The minimum measurable pressure could easily be reduced by using a faster initial rotation velocity, which is limited only by the bandwidth of the driving electronics in the current apparatus.

A second method measures the equilibrium phase lag of the electric dipole moment relative to a driving field, which is induced by the gas drag. This method is used to cross-check the microsphere spinning-rotor gauge against a capacitance manometer. The same bandwidth limitations in the driving electronics limit the minimum measurable pressure of this method to ${\gtrsim}10^{-5}~$mbar.

A geometric calibration factor $\kappa$ related to the microsphere radius is independently determined by measuring the torsional drag as a function of gas species, since the drag depends on the momentum imparted by gas particle collisions and, thus, the molecular mass. The torsional drag $\beta_{\rm rot}$ has been shown to scale inversely to the thermal velocity and, thus, $\beta_{\rm rot} \propto m_0^{1/2}$, with $m_0$ being the molecular mass, validating the rotational dynamics presented here and in Refs.~\onlinecite{Cavalleri:2010,Martinetz:2018}, and opening the possibility to measure the effective molecular mass of a mixture of gases without ionization and directly in moderate vacuum.

\vspace*{-0.3cm}
\begin{acknowledgments}
\vspace*{-0.3cm}

The authors would like to thank the Moore group (Yale) for general discussions related to trapping microspheres. This work was supported, in part, by the National Science Foundation (NSF) under Grant No.~PHY1802952, the Office of Naval Research (ONR) under Grant No.~N00014-18-1-2409, and the Heising-Simons Foundation.  A.K. acknowledges the partial support of a William~M. and Jane~D. Fairbank Postdoctoral Fellowship of Stanford University. N.P. acknowledges the partial support of the Koret Foundation. 

\end{acknowledgments}

\bibliographystyle{apsrev4-1}
\bibliography{spinning_rotor_gauge}

\end{document}